\documentclass[11pt]{article}

\usepackage{authblk}
\usepackage{graphicx}%
\usepackage{multirow}%
\usepackage{amsmath,amssymb,amsfonts}%
\usepackage{amsthm}%
\usepackage[title]{appendix}%
\usepackage{xcolor}%
\usepackage{textcomp}%
\usepackage{booktabs}%
\usepackage{algorithm}%
\usepackage{algorithmicx}%
\usepackage{algpseudocode}%

\providecommand{\keywords}[1]{%
  \par\vspace{0.5em}\noindent\textbf{Keywords:} #1\par
}

\begin{document}

\title{Microscopic study of low-lying energy levels and electromagnetic properties in even-even $^{168-178}$Yb nuclei}

\author[1]{Carlos E. Vargas\thanks{\texttt{cavargas@uv.mx}}}

\author[2]{V\'ictor Vel\'azquez-Aguilar}

\affil[1]{Facultad de F\'isica, Universidad Veracruzana, Paseo No. 112, Desarrollo Hab. Nvo. Xalapa, Xalapa, 91097, Veracruz, Mexico}

\affil[2]{Facultad de Ciencias, Universidad Nacional Aut\'onoma de M\'exico, Apartado Postal 70-542, Mexico City,
04510, Mexico City, M\'exico}

\maketitle

\begin{abstract}Employing methods based on symmetries for the theoretical description of rare-earth nuclei offers many advantages. Among these, the pseudo-SU(3) shell model has proved to be a very useful method to describe characteristics of these systems and to understand various properties. A theoretical description of the low-lying energy spectrum and electromagnetic properties of isotopes $^{168-178}$Yb is carried out for the first time with this model, comparing theoretical results with the experimental values, where possible. The Hamiltonian includes the $Q \cdot Q$ term which preserves the symmetry, as well as the breaking symmetry of Nilsson and pairing terms. Additionally, three rotor type terms are included that allow us to make a subtle adjustment of the spectrum. The results show that the energy spectrum, the B(E2) transitions, the g-factors and the electric quadrupole moments can be described adequately with the model. Although the model is a powerful tool in the description of low-lying properties of normal parity in heavy deformed nuclei, it finds its strongest limitation in the abnormal parity sector, which has been left out of the description.\end{abstract}


\section{Introduction}\label{sec1}

The study of the atomic nuclei is a difficult task to perform. It is a quantum finite system formed by many bodies with fermionic characteristics that interact via a complex force. The construction of a theoretical model aimed to study the nucleus should consider these characteristics, in a way that it would be able to model the complex nucleus dynamic and contribute to interpreting the experimental results in a consistent way.

The low-lying energy spectrum can be reasonably well described with nuclear models very different from each other. Nevertheless, there are other observables whose adequate description represent a major challenge for the theoretical models. Electromagnetic transitions between two states of a nucleus constitute one of these tests, however, such observables depend on the wave function of the initial ($|i\rangle$) and final ($|f\rangle$) many-body states, which might be partially incorrect but compensate one another. Observables that involve only one state constitute a more consistent proof for the model. Some examples of such quantities are the magnetic dipole moment, which give information on the microscopic structure, and the electric quadrupole moment, which allow us to determine whether a nucleus is prolate or oblate \cite{Ham09}. Two important aspects that determine the nuclear model to be used are the valence space associated to heavy nuclei and the Hamiltonian used. A very elegant and efficient way to avoid the problem associated with a very large valence space is to use symmetries, which allows us to perform a systematic and appropriate truncation of the Hilbert space \cite{Cau05}. On the other hand, the use of a realistic Hamiltonian would allow us to include the main terms of the nuclear interaction \cite{Rin79}.

As mentioned in the previous paragraph, the use of symmetries in the shell model would allow us to make a simple and elegant description of the heavy nuclei, avoiding the problem associated with the oversized valence space. The symmetry inspired SU(3) shell-model \cite{Ell58,Har68}, has been used to study light nuclei ($A<$50), where the mean field of the harmonic oscillator and a residual interaction of the quadrupole-quadrupole type permits the description of the predominant characteristics of the nuclear spectrum. In heavier nuclei, the spin-orbit interaction breaks this symmetry so the model stops being useful. In a surprising way, the pseudo-spin symmetry \cite{Gin99} appears at the same time due to the movement of the nucleons in a mean relativistic field, giving rise to a pseudo-SU(3) scheme \cite{Hec69,Ari69}. This symmetry refers to the fact that orbitals of independent particle with $j=l-1/2$ and $j=(l-2)+1/2$ in shell $\eta$ are found quasi-degenerate, thus, they can be labeled as pseudo-spin doublets with quantum numbers $\tilde{j}=j$, $\tilde{\eta}=\eta-1$ and $\tilde{l}=l-1$ \cite{Rat73}.

On the first applications of the pseudo-SU(3) model, pseudo-spin symmetry was used as a dynamic symmetry \cite{Cas87}, which allowed the study of the M1 and E2 transitions towards the ground state $0^+$ in rare earth nuclei and actinides. Later, the development of sophisticated computer codes \cite{Bah94} has enabled to calculate the matrix elements of physical operators which induce the breaking of the symmetry of pseudo-spin. This implies the mixture of representations of different pseudo-SU(3) irreps. With this tool on hand, it has been possible to perform various microscopic studies on rare earth nuclei, such as the description of normal parity bands, E2 and M1 in even-even heavy deformed nuclei \cite{Pop00,Dra01}, the systematics in the structure of low-lying, nonyrast bandhead configurations \cite{Pop04} or the predominance between the orbital and spin contribution to the scissors type oscillation mode in odd-mass rare-earth nuclei \cite{Var03}. More recently, the scheme has been used to describe the low energy spectrum and the electromagnetic properties in the isotope chain $^{160-170}$Dy \cite{Var13,Var17}.

Most of the deformed nuclei in the rare earth region of the nuclear landscape show a quadrupolar prolate type deformation (cigar-like shape) \cite{Rob09} and the transition from prolate to oblate shape only appears for nuclei with a mass number ($A$) around 190. This way, the stable isotopes of Yb that are studied in the present work (those with $168<A<178$) have a prolate deformation. These nuclei have a large number of protons and neutrons of valence, so it is of great interest to study their low energy collective nuclear properties. For example the energy of the state $2^+$ is minimal for the isotope with N=104 ($^{174}$Yb, $E_{2^+}=76.47$eV), while the quadrupole deformation parameter takes its maximum value for N=102 ($^{172}$Yb, $\beta_2=0.329$). The evolution of the deformation of the ground state in this region has been studied using a triaxial meanfield \cite{Rob09}. The conclusions were that for neutron-rich nuclei in Yb ($A>176$), a transition from the prolate shape to the triaxial towards N $\sim 116$ is found, and for even bigger neutron numbers, a potential energy surface with an oblate minimum is obtained. Evolution of the shape in that region of the nuclear landscape requires considering the triaxial degree of freedom \cite{Ots23}. Conducting an analysis of the correlation coefficient has shown that the symmetry of pseudospin is left practically intact when passing from prolate or oblate deformation to triaxial, so for reasonable deformations \cite{Beu97} the pseudo-SU(3) shell model can be used. However, the chain studied in the present contribution ($98 \leq N \leq 108$) is in the region of stability, where deformation is prolate type.

In this article, we present for the first time a study with the pseudo-SU(3) model for the even-even isotope chain $^{168-178}$Yb. The Hamiltonian employed (Eq. \ref{eq:h0}) includes the most relevant terms. Among these is the $Q \cdot Q$ which preserves the pseudo-SU(3) symmetry and the Nilsson and pairing terms that break it, parameterized systematically depending on the mass ($A$) according to the prescriptions widely studied beforehand \cite{Rin79,Duf96}. Additionally, three rotor type terms ($a~K_J^2$, b$J^2$ and $c~\tilde{C_3}$) are used (Eq. \ref{eq:ham}) which allow us to make a finer adjustment of the inertia moment of the rotational bands, as well as the energy of the head of $\gamma$ and $\beta$-bands \cite{Var00,Var04}. The term proportional to $K_J^2$ breaks the SU(3) degeneracy of the different $K$ bands \cite{Naq90}. Using such tools, the low energy structure is studied along the isotopic chain of Yb ($A$=70) starting at $N$=98 and ending with $N$=108. Moreover, results for the intra- and inter-band B(E2) transitions are presented, as well as for the g-factors and the quadrupole moments. 
 
In section \ref{model} a brief description of the pseudo-SU(3) classification scheme and the Hamiltonian employed are discussed. In section \ref{energies}, we present results for energies of levels for ground, $\gamma$ and $\beta$ bands in $^{168}$Yb, $^{170}$Yb, $^{172}$Yb, $^{174}$Yb, $^{176}$Yb, and $^{178}$Yb nuclei. Intra- and inter-band B(E2) transitions, together to the g-factors and electric quadrupole moments in these nuclei are presented in section \ref{sec-be2} and \ref{em-prop}, respectively. Finally, a brief conclusion is given in section \ref{conclu}.

\section{The model}\label{model}

The first step in any application of the pseudo SU(3) model is the construction of the many-body basis. The levels of independent protons and neutrons should fill up from less to more energy for a given deformation, taking the most likely occupations of the orbitals of normal and abnormal parity \cite{Var00}. In Table \ref{occup} the occupation numbers for assigned neutrons for each nucleus are presented. The Yb isotopes have 70 protons, from which 12 fall in normal parity orbitals and 8 in abnormal parity orbitals. As has been done in the calculations made in all cases with the pseudo-SU(3) model to date, the intruder level with opposite parity in each major shell is removed from active consideration and pseudo-orbital and pseudo-spin quantum numbers are assigned to the remaining single-particle states. Nucleons in abnormal parity orbitals are considered to re-normalize the dynamics described using only nucleons in normal parity states, which is reflected, for example, by the use of effective charges greater than those used in typical shell-model calculations for light nuclei. While this has been shown to be a reasonable approach \cite{Cas87}, it is nonetheless a strong assumption and the most important limitation of the present model.

\begin{table}
\begin{tabular}{ccccc}
&&&& \\ \hline \hline
     Nucleus        & $\epsilon_2$ & $n_\nu$ & $n_\nu^N$ & $n_\nu^A$ \\ \hline 
 $^{168}$Yb$_{98}$  &  0.258   &   16    &   10      &  6        \\
 $^{170}$Yb$_{100}$ &  0.267   &   18    &   12      &  6        \\
 $^{172}$Yb$_{102}$ &  0.267   &   20    &   12      &  8        \\
 $^{174}$Yb$_{104}$ &  0.258   &   22    &   14      &  8        \\
 $^{176}$Yb$_{106}$ &  0.250   &   24    &   16      &  8        \\
 $^{178}$Yb$_{108}$ &  0.250   &   26    &   16      & 10        \\ \hline \hline
\end{tabular}
\caption{Deformation ($\epsilon_2$) \cite{Mol95} and occupation numbers ($n$) for neutrons ($\nu$). The 
superscript $N$ and $A$ indicate normal and abnormal parity levels, respectively.}
\label{occup}
\end{table}

Many-particle states of $n_\alpha$ active nucleons ($\alpha = p, n$) in a given (N) normal parity shell $\eta^N_\alpha$ are classified by the following group chain \cite{Cas87,Var00}:

\begin{eqnarray}
~ \{ 1^{n^{N}_\alpha} \} ~~~~~~~ \{ \tilde{f}_\alpha \} ~~~\{ f_\alpha
\} ~\gamma_\alpha ~~~ (\lambda_\alpha , \mu_\alpha ) ~~~ \tilde{S}_\alpha
~~ K_\alpha  \nonumber \\
U(\Omega^N_\alpha ) \supset U(\Omega^N_\alpha / 2 ) \times U(2) \supset
SU(3) \times SU(2) \supset \nonumber \\
\tilde{L}_\alpha  ~~~~~~~~~~~~~~~~~~~~~ J_\alpha ~~~~ \nonumber \\
SO(3) \times SU(2) \supset SU_J(2),
\label{eq:chains}
\end{eqnarray}

\noindent where above each group the quantum numbers that characterize its irreducible representations (irreps) are given. $\gamma_\alpha$ and $K_\alpha$ are multiplicity labels of the indicated reductions. 

The state $| J_i M \rangle$, where J is the total angular moment, M its projection and $i$ the index that lists the states with the same J and M is shaped as a linear combination

\begin{equation} 
| J_i M \rangle = \sum_\beta C^{Ji}_\beta |\beta JM \rangle \label{wf}
\end{equation}

\noindent of the strong coupled proton-neutron states

\begin{eqnarray}
|\beta JM \rangle & \equiv &
| \{ \tilde{f}_\pi \} (\lambda_\pi \mu_\pi) \tilde{S}_\pi, \{ \tilde{f}_\nu \}  
(\lambda_\nu \mu_\nu) \tilde{S}_\nu ; \rho (\lambda \mu ) \kappa \tilde{L},\tilde{S}
JM \rangle~~~~~~~~~~~~~~~~~~~~~~~~~~~~ \nonumber \\
 & = & \sum_{M_L M_S} (\tilde{L} M_L, \tilde{S} M_S | J M ) \sum_{M_{S \pi} M_{S \nu}}
(\tilde{S}_\pi M_{S \pi}, \tilde{S}_\nu M_{S \nu} | \tilde{S} M_S)  \nonumber \\
& & \sum_{k_\pi \kappa_\nu \tilde{L}_\pi \tilde{L}_\nu M_\pi M_\nu }
{ \langle (\lambda_\pi \mu_\pi) \kappa_\pi \tilde{L}_\pi M_\pi ;
(\lambda_\nu \mu_\nu) \kappa_\nu \tilde{L}_\nu M_\nu |
(\lambda \mu ) \kappa \tilde{L} M \rangle}_\rho \label{basis} \\
& & 
| \{ \tilde{f}_\pi \} (\lambda_\pi \mu_\pi) \kappa_\pi \tilde{L}_\pi M_\pi, 
\tilde{S}_\pi M_{S \pi}  \rangle
| \{ \tilde{f}_\nu \} (\lambda_\nu \mu_\nu) \kappa_\nu \tilde{L}_\nu M_\nu, 
\tilde{S}_\nu M_{S \nu} \rangle  ,\nonumber
\end{eqnarray}
where $(-,-|-)$ and $\langle-;-|- \rangle$ are the SU(2) and SU(3) Clebsch Gordan coefficients, respectively, and $\beta$ represents all the summation indices in the expansion of the wavefunction.

The first calculations made with the pseudo-SU(3) model \cite{Cas87} considered only those terms of the Hamiltonian formed by generators of the SU(3) group, so this was considered a dynamical symmetry. With the development of computer codes \cite{Bah94} which allow calculating matrix elements of operators that break the symmetry (such as Nilsson or pairing terms) it has been possible to describe a wide variety of nuclear phenomena. Among some of the applications of this model are the description of the low energy spectrum on the isotopes $^{160-170}$Dy \cite{Var13} chain, and the study of their electromagnetic properties, including the intra- and inter-band B(E2) transitions, as well as the B(M1) values, the electric quadrupole moments and the g-factors  \cite{Var17}.

The pseudo-SU(3) model is a powerful theory that allows us to use a relatively simple truncation scheme based on physical arguments. It is important to note that the configuration space used is made up of pseudo-spin zero and one for protons and neutrons, given that these are the couplings that have most physical relevance. Spin excitations greater than one for proton or neutron have not been included, as it has been determined experimentally that those spin-flip modes are relevant in describing higher energy states than those values considered in this analysis. 

The principal part of the Hamiltonian ($H_0$) contains spherical Nilsson single-particle terms for the protons and neutrons ($H_{sp,\pi[\nu]}$), the pairing ($H_{pair,\pi[\nu]}$) and quadrupole-quadrupole ($\tilde Q \cdot \tilde Q$) interactions parametrized systematically, 
\begin{eqnarray}
H_0 & = & \sum_{\alpha=\pi,\nu} \{ H_{sp,\alpha} - G_\alpha 
~H_{pair,\alpha} \} - \frac{1}{2}~  \chi~ \tilde Q \cdot \tilde Q ~.
\label{eq:h0}
\end{eqnarray}
Additionally, three ‘rotor-like’ terms ($K^2$, $J^2$ and $C_3$) that are diagonal in the SU(3) basis {\bf ,} have been considered:
\begin{eqnarray}
    H & = & H_0 + ~a~ K_J^2~ + ~b~ J^2~ + ~c~ \tilde C_3. \label{eq:ham}
\end{eqnarray}

\noindent The main part of this Hamiltonian (eq. \ref{eq:h0}) contains the essential elements of a realistic interaction that allow the description of the deformed nuclei: The levels of single particle of Nilsson, the pairing correlations, and the quadrupole interaction term. These terms and their parameterization have been studied in depth \cite{Rin79,Duf96,Var00}, which has allowed its intensity to be fixed according to the mass ($A$), so they are not free parameters of the model. 

The single-particle terms ($H_{sp,\alpha}$) have the form:
\begin{eqnarray}
H_{sp,\alpha} = \sum_{i_\alpha} \left[ \hbar \omega_0 \left( \hat{\eta}_{i_\alpha}+\frac{3}{2} \right)
+ C_\alpha {\bf  l}_{i_\alpha} \cdot {\bf s}_{i_\alpha} + D_\alpha {\bf  l}^2_{i_\alpha}\right],    
~~~~~\alpha=\pi,\nu,
\end{eqnarray}
where $C_\alpha$ and $D_\alpha$ are fixed following the usual prescriptions \cite{Rin79}. In the pseudospin basis the spin-orbit and orbit-orbit contributions are small, but they still generate most of the mixing between pseudo SU(3) irreps, in addition to that produced by the pairing term which also breaks the symmetry.

The rotorlike terms in Hamiltonian (\ref{eq:ham}) are used to fine tune the spectra. Their three parameters $a$, $b$, and $c$ have been fixed following the prescriptions given in Ref. \cite{Var00}, where a detailed analysis of each term can be found. Only these three terms are taken as free parameters of the model. In Table \ref{parame} the parameters used in the Hamiltonian (\ref{eq:ham}) are shown.

\begin{table}
\begin{tabular}{ccccccc}\hline \hline
Parameter& $^{168}$Yb & $^{170}$Yb & $^{172}$Yb &$^{174}$Yb & $^{176}$Yb & $^{178}$Yb \\ \hline
$\chi$ 	&   6.842     &   6.709    &   6.579    &   6.454   &   6.332    &   6.214    \\
G$_\pi$	& 125.0       & 123.5      & 122.1      & 120.7     & 119.3      & 117.9      \\
G$_\nu$ & 101.2       & 100.0      &  98.8      &  97.7     &  96.6      &  95.5      \\
$a$	&  90.0       & 220.0      & 300.0      & 900.0     & 220.0      & 220.0      \\   
$b$	&   2.5       &   1.0      &   0.6      &  -6.0     &  -6.0      &  -6.0      \\
$c$	& 0.0180      & 0.0100     & 0.0100     & 0.0249    & 0.0165     & 0.0165     \\ \hline \hline
\end{tabular}
\caption{Parameters (in keV) used in the Hamiltonian (\ref{eq:ham}).}
\label{parame}\end{table}

To clarify the effect of these rotorlike terms upon the energies, we may take as an example the $b J^2$ term. Its effect is to give an additional moment of inertia to the rotational bands, helping to diminish the energy of rotational states. When this term is considered, the energy of the $J_{band}^\pi=2_{g.s.b.}^+$ states in the ytterbium isotopes studied in the present contribution reduces its value on average 34 keV. In the specific case of $^{174}$Yb, the energy of the $J^\pi=2_{g.s.b.}^+$ state changes from 111 keV (with $b=0$) to 77 keV (with $b=-6.0$ keV). In $^{176}$Yb it changes from 118 keV ($b=0$) to 82 keV ($b=-6.0$ keV). Those states with higher angular momentum are also affected by this rotor term with even larger changes. Using $^{176}$Yb as example, the state $J^\pi=8_{\gamma}^+$ changes from 1352 keV when $b=0$ to 966 keV when $b=-6.0$ keV. It is important to mention that the rotor terms have influence on the level spacings within the rotational bands, but the wave function holds practically constant \cite{Var00} within 0.001 \%.

In the calculation of B(E2) transitions, the g-factors and the quadrupole moments, the wave function obtained on the eigenvalues problem, discussed in this section, has been used. Following the notation of the reference \cite{Cas87}, the  magnetic dipole transition operator is given by 
\begin{eqnarray}
T_{1\mu}(M1) \equiv \mu_N \left( g_\pi^o L_\mu^\pi + g_\nu^o L_\mu^\nu 
+  g_\pi^s S_\mu^\pi + g_\nu^s S_\mu^\nu \right)
\label{T1}
\end{eqnarray}
where $L_\mu^\sigma$ ($S_\mu^\sigma$) and $g^o_\sigma$ ($g^s_\sigma$) are, the orbital (spin) angular momentum and orbital (spin) gyro-magnetic factor for protons ($\sigma = \pi$) and neutrons ($\sigma = \nu$), and $\mu_N$ denotes the nuclear magneton. In our calculation, the orbital and quenched spin g factors (by a factor of 0.7 due to core-polarization effects) for protons and neutrons are used: $g^o_\pi = 1, g^o_\nu = 0, g^s_\pi = (0.7)5.5857$ and $g^s_\nu = -(0.7)3.8263$. The same values are used for all $g$-factor calculations in the present paper, as in the previous pseudo SU(3) shell model calculations, without any adjustment for individual nuclei. It is important to highlight that in the present calculation the configuration space was built considering the states with $\tilde{S}_{\pi,\nu} = 0$ and $1$, in a way that the matrix elements of the spin operators of proton and neutron have a significant contribution.

In a similar way, the operator of the electric quadrupole transition is defined as
\begin{eqnarray}
T_{2\mu}(E2) & \equiv & b_0^2 \left( e_\pi \sum_{i_\pi} \sqrt{\frac{16\pi}{5}} r_\pi^2(i) 
Y_{2\mu}(\theta_i^\pi,\phi_i^\pi) \right) \nonumber \\
&  & + b_0^2 \left( e_\nu \sum_{i_\nu} \sqrt{\frac{16\pi}{5}} r_\nu^2(i) 
Y_{2\mu}(\theta_i^\nu,\phi_i^\nu) \right)\nonumber,
\end{eqnarray}
where $b_0 = A^{1/3}$ fm is the harmonic-oscillator size parameter \cite{Tro95}.

The expression for reduced matrix elements of these tensor operators, just as the reduced probabilities for magnetic dipole and electric quadrupole radiation, can be found at Ref. \cite{Var17}. As the transition operator E2 is made up by generators of the SU(3) group, transitions cannot be predicted between different irreps, so at the limit of the dynamic symmetry of the model, quadrupole electric transitions between states belonging to the ground band and the first excited band $K^\pi=0^+$ are forbidden. In a more realistic way, when including the breaking symmetry terms of Nilsson or pairing, such transitions are allowed, their values being much lower than transitions within the rotational band. 

\section{Low-lying energy spectra}\label{energies}

Starting from the base states and the Hamiltonian discussed in the previous section, the results for the energy of the levels on $^{168-178}$Yb isotopes are presented. 
\begin{figure}[h]%
\centering
\includegraphics[width=0.9\textwidth]{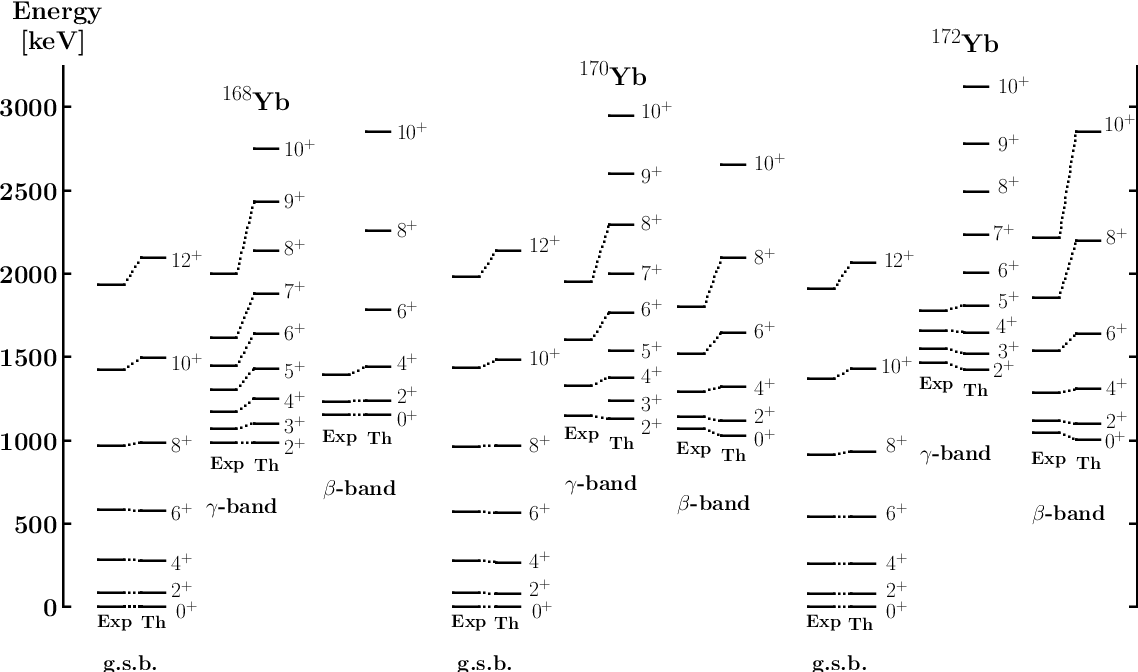}
\caption{Experimental and theoretical energies (in keV) of ground, $\gamma$ and $\beta$-bands in $^{168-172}$Yb nuclei. The labels indicate the total angular momentum and parity of each level. Experimental data \cite{Cor10,Bag18,Sin95} are plotted on the left-hand side of each column and theoretical ones on the right-hand side. The correspondence between theoretical and experimental levels is indicated by dotted-lines.}\label{energies1}
\end{figure}
Fig. \ref{energies1} and \ref{energies2} show a comparison between the experimental and theoretical energies for the ground state, $\gamma$ and $\beta$-bands in the isotope chains $^{168-172}$Yb and $^{174-178}$Yb, respectively. For the six studied nuclei, the energies of the states from the ground band up to $J^\pi=12^+$ have been calculated with the model.

For the $\gamma$-bands, the model predicts a very small moment of inertia in $^{168-172}$Yb, while in the $^{174-178}$Yb isotopes, the predicted moment of inertia is in accordance with the experimental values. For the $\beta$ bands, the theoretical moment of inertia is very small for all the cases, so the predicted theoretical energies for the states with a large angular momentum are much higher than the values observed experimentally, as can be noted on Figures \ref{energies1} and \ref{energies2}. For example, in $^{172}$Yb, the $J^\pi=10^+$ state of the $\beta$ band, an energy of 2853 keV is predicted, which is 641 keV above the experimental value. Otherwise, it is interesting to notice that in the two heavier nuclei studied in this work ($^{176-178}$Yb), the theoretical $\beta$ bands have an aspect of vibrational bands with equidistant energy levels.

It is interesting to discuss the change in the values for $b$ parameter of the rotor type term ($J^2$) in Table \ref{parame} when passing from $^{168}$Yb to $^{178}$Yb and the effect it has over the moment of inertia of the rotational bands. In  $^{168}$Yb the $b$ parameter of the Hamiltonian (eq. \ref{eq:ham}) takes the value 2.5 keV and for the heavier isotopes this is decreasing to 1 keV, then to 0.6 keV until it reaches the $b$ = -6 keV value in $^{174-178}$Yb. This term contibutes to diminish (for values $b>0$) the moment of inertia of the rotational bands. In other words, the rotor type term $bJ^2$ in the Hamiltonian (eq. \ref{eq:ham}) introduces a slight correction in the moment of inertia obtained with the Hamiltonian (\ref{eq:h0}), as described in the previous section. For example, in $^{170}$Yb the energy obtained for the $J^\pi=8^+$ state of the ground band with the Hamiltonian $H_0$ (eq. \ref{eq:h0}) is 875 keV, but when taking $b=1$ keV in the Hamiltonian $H$ (eq. \ref{eq:ham}), the predicted theoretical value is 965 keV, which is very close to the experimental value (963 keV). The pseudo-SU(3) model with the Hamiltonian (\ref{eq:h0}) predicts a strong collectivity as is implied by the data in $^{168-172}$Yb, but a weak collectivity in $^{174-178}$Yb isotopes. This phenomenology is corrected in the model when entering the parameter $bJ^2$ in the eq. (\ref{eq:ham}). The value of the $b$ parameter used, goes from 2.5 keV to -6 keV, indicating that the model predicts a large moment of inertia in the isotopes $^{168}$Yb and $^{170}$Yb, while in the  $^{174-178}$Yb nuclei, the model requires an additional moment of inertia that is provided by the term $bJ^2$ in the eq. (\ref{eq:ham}). It should be pointed out that the absence of nucleons in the intruder sector could help to explain the loss of collectivity in $^{174-178}$Yb, but at this stage of calculations we are not able to include the intruder sector, presenting a challenge for future works with the model.

Referring to the other two rotor type parameters ($a$ and $c$) in the Hamiltonian (\ref{eq:ham}), these entered corrections in the energy of the band head of $\gamma$ and $\beta$-bands, respectively. The particular effect of each of these rotor type terms have been studied extensively beforehand \cite{Var00}.
\begin{figure}[h]
\centering
\includegraphics[width=0.9\textwidth]{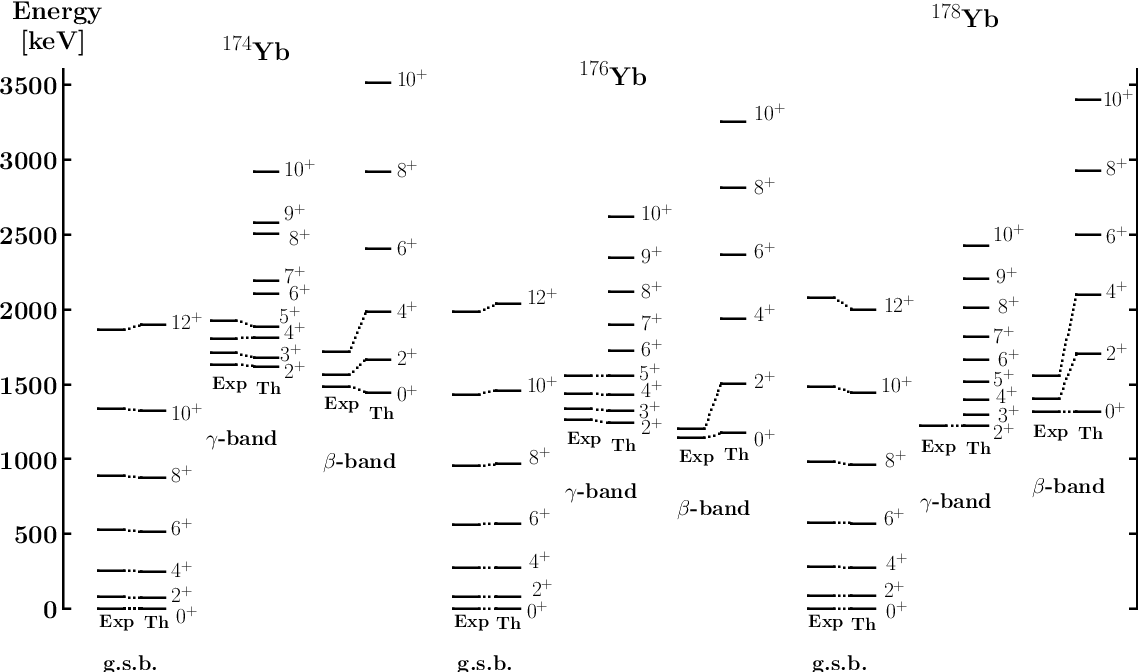}
\caption{Experimental \cite{Bro99,Bas06,Ach09} and theoretical energies (in keV) in $^{174-178}$Yb nuclei. The labels are the same as in Figure \ref{energies1}.}\label{energies2}
\end{figure}
Through the coupling of the $\tilde{S}_\pi$ and $\tilde{S}_\nu$ components it is possible to build the total pseudo-spin content of the nuclear wave function. As has been the case in previous works with the model, the ground state bands are composed predominantly by $\tilde{S}=0$ (these are the functions having maximum orbital symmetry), with very small mixing of $\tilde{S}=1$, which have less orbital symmetry \cite{Har68}. For example, in the ground state band in $^{168}$Yb, the wave function of the $0^+$ state has 10$\%$ of the $\tilde{S}=1$ component, but for the $10^+$ state of the same band, the $\tilde{S}=1$ content is almost completely absent. Nevertheless, the $\gamma$-bands have larger components of $\tilde{S}=1$; taking $^{168}$Yb again as example, the $2^+$ band-head has 97$\%$ of $\tilde{S}=1$ content, but the $10^+$ of the same band has only 51$\%$ of $\tilde{S}=1$. Large components of $\tilde{S}=1$ are also observed in $\beta$-bands. These results show the importance of the $\tilde{S}=1$ contribution in the description of excited $\gamma$ and $\beta$-bands, as it has been pointed out in Ref. \cite{Var13}.

\section{B(E2) transition strengths}\label{sec-be2}

In this contribution we use the pseudo-SU(3) model to investigate the electromagnetic properties of the $^{168-178}$Yb nuclei. Calculated B(E2) values are given in units of $e^2 b^2 \times 10^{-2}$. The effective charges used in the electric quadrupole operator $Q_\mu$ \cite{Var17} are $e_\pi = 2.3$ and $e_\nu=1.3$. These values are the same used in the pseudo-SU(3) studies up to now allowing the description of both intra- and inter-band B(E2)s \cite{Var01}. They are larger than those used in standard calculations of quadrupole transitions \cite{Rin79} due to the absence of nucleons in intruder levels. In order to follow the same guidelines of previous works, we have not modified the effective charges. Smaller effective charges might improve the agreement with experiment for intra-band E2 strengths as well as the quadrupole moments, but that calculation remains as future work. The values used for the effective charges are fixed and not considered free parameters of the model. Table \ref{be2} shows the intra-band B(E2) transitions for the states of the ground, $\gamma$ and $\beta$-bands. As can be seen in this Table, values of the B(E2) are overestimated compared to available experimental data \cite{Cor10,Bag18,Sin95,Bro99,Bas06}. This is because the model predicts highly collective states and with very small variations in the wave function from one state to another of the same band, so there is a very large overlap of these functions, producing extremely high values of B(E2). The $\gamma$ and $\beta$-bands also show high collectivity, with B(E2) transitions larger than 100 $e^2 b^2 \times 10^{-2}$ in almost every case. It is interesting to mention the variations that the model predicts for intra-band B(E2) strengths depending on the number of neutrons. For the ground state band, the maximum values found are for the $^{172}$Yb nuclei ($N$=102). For the $\beta$-band, the maximum values for the transitions $0_\beta^+ \rightarrow 2_\beta^+$ and $2_\beta^+ \rightarrow 4_\beta^+$ occur also in $^{172}$Yb ($N$=102), but for transitions with $J^+_i \geq 4^+$ the maximum values occur for $^{178}$Yb ($N$=108). On the $\gamma$-band, the maximum values are those presented in the $^{172}$Yb and $^{174}$Yb nuclei.

\begin{table}\hspace{-3.2cm}
\begin{tabular}{c|cc|cc|cc|cc|cc|c}\hline \hline
 & \multicolumn{11}{c}{B(E2) [$e^2b^2 \times 10^{-2}$]}\\ 
$J_{i,band}^{\pi} \rightarrow J_{f,band}^{\pi}$ & \multicolumn{2}{c|}{$^{168}$Yb} & \multicolumn{2}{c|}{$^{170}$Yb} & \multicolumn{2}{c|}{$^{172}$Yb} & \multicolumn{2}{c|}{$^{174}$Yb} & \multicolumn{2}{c|}{$^{176}$Yb} & $^{178}$Yb \\
                 		     & Exp.&Theo.&   Exp.    &Theo.&     Exp.   &Theo.&    Exp.    &Theo.&    Exp.    &Theo.& Theo.\\ \hline
$0^+_{g.s.b.}\rightarrow2^+_{g.s.b.}$& 575$\pm$19 &653& 562$\pm$17 &721 &  602$\pm$6 & 726 & 579$\pm$20 & 721 & 536$\pm$20 & 697 &  702 \\
$2^+_{g.s.b.}  \rightarrow 4^+_{g.s.b.}$    & &  335 & 	     & 369 & 171$\pm$11 & 372 &  61$\pm$5  & 368 & 158$\pm$15 & 357 &  359 \\ 
$4^+_{g.s.b.}  \rightarrow 6^+_{g.s.b.}$    & &  293 &  	     & 323 & 181$\pm$17 & 325 & 213$\pm$29 & 321 & 174$\pm$13 & 313 &  315 \\
$6^+_{g.s.b.}  \rightarrow 8^+_{g.s.b.}$    & &  274 & 201$\pm$17 & 301 & 227$\pm$23 & 303 & 224$\pm$12 & 298 & 176$\pm$29 & 293 &  295 \\
$8^+_{g.s.b.}  \rightarrow 10^+_{g.s.b.}$   & &  261 & 199$\pm$14 & 286 & 213$\pm$13 & 288 & 193$\pm$13 & 282 & 187$\pm$18 & 280 &  283 \\ 
$10^+_{g.s.b.} \rightarrow 12^+_{g.s.b.}$   & &  247 & 150$\pm$12 & 268 & 244$\pm$34 & 271 & 213$\pm$13 & 237 & 181$\pm$23 & 269 &  272 \\\hline
$0^+_{\beta} \rightarrow 2^+_{\beta}$  & &  502 & 	     & 532 &      	& 697 &      	   & 498 & 	      & 500 &  505 \\
$2^+_{\beta} \rightarrow 4^+_{\beta}$  & &  282 & 	     & 292 &      	& 356 &      	   & 262 & 	      & 289 &  296 \\
$4^+_{\beta} \rightarrow 6^+_{\beta}$  & &  244 & 	     & 291 &      	& 311 &      	   & 241 & 	      & 305 &  314 \\
$6^+_{\beta} \rightarrow 8^+_{\beta}$  & &  208 & 	     & 283 &      	& 290 &      	   & 241 & 	      & 316 &  319 \\
$8^+_{\beta} \rightarrow 10^+_{\beta}$ & &  197 & 	     & 273 &      	& 276 &      	   & 248 & 	      & 316 &  318 \\
$10^+_{\beta}\rightarrow 12^+_{\beta}$ & &  181 & 	     & 207 &      	& 163 &      	   & 257 & 	      &     &  314 \\\hline
$2^+_{\gamma}\rightarrow 3^+_{\gamma}$ & &  288 & 	     & 261 &      	& 349 &      	   & 141 & 	      & 348 &  350 \\
$3^+_{\gamma}\rightarrow 4^+_{\gamma}$ & &  198 & 	     & 150 &      	& 238 &      	   &  72 & 	      & 235 &  237 \\
$4^+_{\gamma}\rightarrow 5^+_{\gamma}$ & &  127 & 	     & 104 &      	& 161 &      	   &  45 & 	      & 161 &  163 \\
$5^+_{\gamma}\rightarrow 6^+_{\gamma}$ & &   73 & 	     &  66 &      	& 113 &      	   &  29 & 	      & 110 &  111 \\
$6^+_{\gamma}\rightarrow 7^+_{\gamma}$ & &   24 & 	     &  48 &      	&  83 &      	   &  21 & 	      &  84 &   85 \\
$7^+_{\gamma}\rightarrow 8^+_{\gamma}$ & &   17 & 	     &  34 &      	&  62 &      	   &  15 & 	      &  59 &   59 \\
$8^+_{\gamma}\rightarrow 9^+_{\gamma}$ & &   22 & 	     &  26 &      	&  48 &      	   &  11 & 	      &  50 &   50 \\
$9^+_{\gamma}\rightarrow 10^+_{\gamma}$& &   15 & 	     &  20 &      	&  37 &      	   &   9 & 	      &  32 &   33 \\\hline
$2^+_{\gamma}\rightarrow 4^+_{\gamma}$ & &  119 & 	     & 189 &      	& 148 &      	   & 305 & 	      & 147 &  148 \\
$3^+_{\gamma}\rightarrow 5^+_{\gamma}$ & &  159 & 	     & 199 &      	& 206 &      	   & 305 & 	      & 206 &  207 \\
$4^+_{\gamma}\rightarrow 6^+_{\gamma}$ & &  160 & 	     & 244 &      	& 230 &      	   & 298 & 	      & 229 &  231 \\
$5^+_{\gamma}\rightarrow 7^+_{\gamma}$ & &   99 & 	     & 237 &      	& 241 &      	   & 294 & 	      & 243 &  245 \\
$6^+_{\gamma}\rightarrow 8^+_{\gamma}$ & &  188 & 	     & 248 &      	& 245 &      	   & 286 & 	      & 242 &  241 \\
$7^+_{\gamma}\rightarrow 9^+_{\gamma}$ & &  110 & 	     & 244 &      	& 247 &      	   & 282 & 	      & 249 &  244 \\
$8^+_{\gamma}\rightarrow 10^+_{\gamma}$& &  193 & 	     & 242 &      	& 245 &      	   & 274 & 	      & 234 &  236 \\
$9^+_{\gamma}\rightarrow 11^+_{\gamma}$& &  204 & 	     & 241 &      	& 243 &      	   & 270 & 	      & 243 &  246 \\ \hline \hline
\end{tabular}
\caption{B(E2;$J^+_i \rightarrow J^+_f$) intra-band transitions in $^{168-178}$Yb nuclei (given in $e^2b^2 \times 10^{-2}$). The first column gives the initial ($J_i$) and final ($J_f$) values of the angular momentum. From second to twelfth column the experimental (Exp.) and pseudo-SU(3) model calculations (Theo.) are shown. Effective charges are $e_\pi$=2.3 and $e_\nu$=1.3.}
\label{be2}\end{table}

Table \ref{inter} reports the inter-band B(E2) strengths between states of ground, $\gamma$ and $\beta$-bands. These values are significantly smaller than those shown in Table \ref{be2}, because the wave functions of states belonging  to different bands have very different components. Nevertheless, there are some B(E2)s with large values, which is the result of a strong overlap between the wave functions of the states. It is interesting to observe that the theoretical values for the inter-band transitions are underestimated regarding the experimental data. In most cases, the predicted values are under the 1.0 $e^2b^2 \times 10^{-2}$. An interesting case occurs in the transitions between states of bands $\gamma$ and the  $\beta$ in $^{170}$Yb. For the transition B(E2; $0^+_\beta \rightarrow 2^+_\gamma$), the intensity is $151.481~e^2b^2 \times 10^{-2}$, a value that is high above those calculated for the inter-band transitions shown in Table \ref{inter}. This results from the overlap of very similar wave functions between the states $0^+_\beta$ and $2^+_\gamma$, and possibly because of the close proximity in energy of these states in $^{170}$Yb. Some clues to the increase in values in the B(E2) strengths could come from the $\gamma$-ray branching ratios, but the relevant transitions have not been determined. Other transitions like B(E2; $2^+_\beta \rightarrow 3^+_\gamma$) or B(E2; $3^+_\gamma \rightarrow 4^+_\beta$) are also very intense, which indicates a big mix of the $\gamma$ and $\beta$-bands in this nucleus. Also noteworthy is the large quantity of inter-band transitions in the $^{174-178}$Yb nuclei, for which the transition values are very small (B(E2) $< 0.001~e^2b^2 \times 10^{-2}$), a fact that reflects little mixing in the components of rotational bands. It is interesting to mention that in the limit of the dynamic symmetry, inter-band transitions are zero, as each band is formed by eigenstates of SU(3) and the overlap between them is null. The presence of symmetry-breaking terms in the Hamiltonian, create a mix of irreps, so when calculating the inter-band B(E2) transitions, some overlaps occur, implying intensity different from zero. 

\begin{table}\hspace{-4.7cm}
\begin{tabular}{c|cc|cc|cc|cc|cc|c}\hline \hline
 & \multicolumn{11}{c}{B(E2) [$e^2b^2 \times 10^{-2}$]}\\ 
$J_{i,band}^{\pi} \rightarrow J_{f,band}^{\pi}$ & \multicolumn{2}{c|}{$^{168}$Yb} & \multicolumn{2}{c|}{$^{170}$Yb} & \multicolumn{2}{c|}{$^{172}$Yb} & \multicolumn{2}{c|}{$^{174}$Yb} & \multicolumn{2}{c|}{$^{176}$Yb} & $^{178}$Yb \\
                 		         & Exp.           &  Theo.  &   Exp.         &  Theo.  &     Exp.      & Theo.   &    Exp.    &Theo.    &    Exp.       & Theo. &  Theo.  \\ \hline
$0^+_{g.s.}  \rightarrow  2^+_{\beta}$   &  4.96$\pm$0.55 & 0.001  &  3.02$\pm$0.59 & 0.041  & 0.68$\pm$0.03 & 0.024  &          &   *     &               &    *    &   *    \\
$0^+_{g.s.}  \rightarrow  2^+_{\gamma}$  & 13.77$\pm$1.93 & 0.009  &  7.55$\pm$1.68 & 0.001  & 3.78$\pm$0.31 & 0.022  &          &  0.766 & 5.27$\pm$0.62 & 11.325 & 11.363  \\
$0^+_{\beta}  \rightarrow  2^+_{g.s.}$   &                &   *    &                & 0.159  & 2.04$\pm$0.56 & 0.171  &0.81$\pm$0.63 & *   &               &  *      &   *    \\
$0^+_{\beta}  \rightarrow  2^+_{\gamma}$ &                & 0.006  &                & 151.481& 6.87$\pm$0.68 & 0.735  &          &   *     &               &  *      &   *    \\
$2^+_{g.s.}  \rightarrow  3^+_{\gamma}$  &                & 0.002  &                & 0.009  &               & 0.011  &          & 0.171 &               & 5.647  & 5.676   \\
$2^+_{g.s.}  \rightarrow  4^+_{\gamma}$  &                &   *    &                & 0.052  & 3.98$\pm$2.84 & 0.002  &          & 0.268 &               & 1.205  & 1.193   \\
$2^+_{g.s.}  \rightarrow  4^+_{\beta}$   &                & 0.006  &                & 0.054  &               & 0.107  &          &  *     &               &  *      &   *    \\
$2^+_{\gamma}  \rightarrow  4^+_{g.s.}$  &  0.99          & 0.002  & 0.27$\pm$0.06  & 0.109  & 0.07$\pm$0.01 & 0.007  &          & 0.205 &               & 0.496  & 0.523   \\
$2^+_{\gamma}  \rightarrow  4^+_{\beta}$ &                &   *    &                & 3.653  &               & 0.377  &          &  *     &               &  *      &   *    \\
$2^+_{\beta}  \rightarrow  3^+_{\gamma}$ &                & 0.366  &                & 73.863 &               & 0.365  &          &  *     &               &  *      &   *    \\
$2^+_{\beta}  \rightarrow  4^+_{g.s.}$   &                &   *    &                & 0.157  & 1.42$\pm$0.17 & 0.275  &          &  *     &               &  *      &   *    \\
$2^+_{\beta}  \rightarrow  4^+_{\gamma}$ &                & 0.019  &                & 14.625 &               & 0.156  &          &   *     &               &  *      &   *    \\
$3^+_{\gamma}  \rightarrow  4^+_{g.s.}$  &                & 0.006  &  & 0.041  &  &   0.033  &  & 0.345  & &3.742   & 3.792  \\
$3^+_{\gamma}  \rightarrow  4^+_{\beta}$ &                & 0.002  &  & 85.137 &  &   2.205  &  &   *    & &  *     &   *    \\
$4^+_{g.s.}  \rightarrow  6^+_{\beta}$   &                &   *    &  & 0.076  &  &   0.182  &  &   *    & &  *     &   *    \\
$4^+_{\gamma}  \rightarrow  6^+_{g.s.}$  &                & 0.001  &  & 0.199  &  &   0.022  &  & 0.208  & &1.439   & 1.460  \\
$4^+_{\gamma}  \rightarrow  6^+_{\beta}$ &                & 0.008  &  & 0.672  &  &   1.117  &  &   *    & &  *     &   *    \\
$4^+_{\beta}  \rightarrow  5^+_{\gamma}$ &                & 0.037  &  & 38.059 &  &     *    &  &   *    & &  *     &   *    \\
$4^+_{\beta}  \rightarrow  6^+_{g.s.}$   &                & 0.005  &  & 0.178  &  &   0.384  &  &   *    & &  *     &   *    \\
$4^+_{\beta}  \rightarrow  6^+_{\gamma}$ &                & 1.287  &  & 1.364  &  &   0.645  &  &   *    & &  *     &   *    \\
$5^+_{\gamma}  \rightarrow  6^+_{g.s.}$  &                &   *    &  & 0.050  &  &   0.045  &  & 0.341  & &5.700   & 5.800  \\
$5^+_{\gamma}  \rightarrow  6^+_{\beta}$ &                & 0.012  &  & 42.549 &  &   3.669  &  &   *    & &  *     &   *    \\ \hline \hline
\end{tabular}
\caption{B(E2;$J^+_i \rightarrow J^+_f$) inter-band transitions in $^{168-178}$Yb nuclei (given in $e^2b^2 \times 10^{-2}$). The first column gives the initial ($J_i$) and final ($J_f$) values of the angular momentum. From second to twelfth column, the experimental (Exp.) and pseudo-SU(3) model calculations (Theo.) are shown. Effective charges are $e_\pi$=2.3 and $e_\nu$=1.3. The symbol * indicates B(E2) $ < 0.001 e^2b^2 \times 10^{-2}$.}
\label{inter}\end{table}

\section{g-factors and quadrupole moments}
\label{em-prop}

Apart from the B(E2) electromagnetic strengths, in this contribution, we used the pseudo-SU(3) model to study the gyromagnetic factors and the electric quadrupole moments in the $^{168-178}$Yb nuclei. In Table \ref{g-fact} gyromagnetic factors are presented. It is interesting to mention some outstanding behaviors. The experimental value of the $g(2_{g.s.}^+)$-factor, has only been measured in the $^{170-176}$Yb isotopes, being almost constant for these nuclei and with values within the range 0.330 and 0.338 $\mu_N$. In contrast, the calculated value using the present model for the $g(2_{g.s.}^+)$ state in $^{168}$Yb is 0.24, and it gradually increases with the number of neutrons until it reaches a 0.37 in $^{178}$Yb. It is important to mention the $^{174}$Yb, in which the theoretical value lies within the error bars of the experimental data. 

For the states with a larger angular momentum of the ground state band, the model predicts a slight increase in the values of the g factor in the $^{168,170,172}$Yb nuclei, while in the $^{174,176,178}$Yb nuclei, the values of the g factor remain constant, as can be seen on Table \ref{g-fact}. Because intruder orbitals have not been considered in the present calculation, the pseudo-su(3) model predicts near constant g factors. This Table also presents the values of the g-factors for the states of $\gamma$ and $\beta$-band, where the change in terms of the angular momentum can be analyzed. In most of the nuclei, we can observe an increase in the value of the g-factor when passing from $J^\pi = 2^+$ to $10^+$, although, there are some exceptions that can be seen also. An outstanding case is that of the  $\gamma$-band in $^{168}$Yb, where the maximum value is $g(2^+_\gamma)=1.11$, decreasing until $g(10^+_\gamma)=0.26$, or $\beta$-band in the same nucleus, where there is a slight reduction. Another exception is seen in $^{172}$Yb, in the $\beta$-band, for which $g=0.32$ for all the sates of the band. Some cases like the $7_\gamma^+$ in $^{170}$Yb or the $6_\gamma^+$ en $^{174}$Yb, the calculated value (0.37) is very close to the reference $Z/A=0.4$ for the state of rotational bands. The g-factors are a sensitive proof of the microscopic structure of individual states \cite{Sun07,Hol01}.

As can be seen in Table \ref{g-fact}, the values of the g factors of the $\gamma$-band differ significantly from those for the ground state band. However, at this moment we cannot offer physical arguments to explain this discrepancy with respect to the standard collective model approach. On the other hand, Alfter et al. \cite{Alf96} have shown that $g(2^+_2)/g(2^+_1) < 1$, as found in many cases reported in Table \ref{g-fact}. It would be interesting to analyze the relationship between the M1 and the g factors with the model to determine the origin of the deviation in $^{168}$Yb. This is out of the scope of this work and is left as a proposal for future works.

\begin{table}\hspace{-2.7cm}
\begin{tabular}{c|c|cc|cc|cc|cc|c}\hline \hline
 & \multicolumn{10}{c}{g-factors $(J^\pi_{band})~[\mu_N]$} \\ 
$J_{band}^{\pi}$ &   $^{168}$Yb & \multicolumn{2}{c}{$^{170}$Yb} & \multicolumn{2}{c}{$^{172}$Yb} & \multicolumn{2}{c}{$^{174}$Yb} & \multicolumn{2}{c}{$^{176}$Yb} & $^{178}$Yb \\
                 &   Theo. &   Exp.      & Theo. &    Exp.     & Theo. &    Exp.     &Theo. &    Exp.     &Theo. & Theo. \\ \hline
$2^+_{g.s.b.}$   & 0.24 &0.337$\pm$0.004& 0.22&0.334$\pm$0.008& 0.25 &0.338$\pm$0.004&0.34&0.33$\pm$0.015& 0.36 &  0.37 \\
$4^+_{g.s.b.}$   & 0.24  &             & 0.24 &0.342$\pm$0.013& 0.25 &             & 0.34 & & 0.35 &  0.35 \\
$6^+_{g.s.b.}$   & 0.24  &             & 0.24  &             & 0.25  &             & 0.34 & & 0.35 &  0.35 \\
$8^+_{g.s.b.}$   & 0.24  &             & 0.25  &             & 0.25  &             & 0.34 & & 0.35 &  0.35 \\
$10^+_{g.s.b.}$  & 0.28  &             & 0.26  &             & 0.26  &             & 0.34 & & 0.35 &  0.35 \\ \hline
$2^+_{\gamma}$   & 1.11  &             & 0.19  &             & 0.11  &             & 0.32 & & 0.11 &  0.11 \\
$3^+_{\gamma}$   & 0.72  &             & 0.23  &             & 0.23  &             & 0.32 & & 0.23 &  0.23 \\
$4^+_{\gamma}$   & 0.55  &             & 0.29  &             & 0.27  &             & 0.36 & & 0.28 &  0.29 \\
$5^+_{\gamma}$   & 0.46  &             & 0.30  &             & 0.30  &             & 0.33 & & 0.31 &  0.31 \\
$6^+_{\gamma}$   & 0.39  &             & 0.32  &             & 0.31  &             & 0.37 & & 0.32 &  0.32 \\
$7^+_{\gamma}$   & 0.29  &             & 0.37  &             & 0.32  &             & 0.34 & & 0.33 &  0.33 \\
$8^+_{\gamma}$   & 0.32  &             & 0.32  &             & 0.32  &             & 0.36 & & 0.34 &  0.34 \\
$9^+_{\gamma}$   & 0.31  &             & 0.33  &             & 0.33  &             & 0.34 & & 0.34 &  0.34 \\
$10^+_{\gamma}$  & 0.26  &             & 0.33  &             & 0.33  &             & 0.36 & & 0.34 &  0.34 \\ \hline
$2^+_{\beta}$    & 0.33  &             & 0.23  &             & 0.32  &             & 0.29 & & 0.25 &  0.25 \\
$4^+_{\beta}$    & 0.31  &             & 0.29  &             & 0.32  &             & 0.29 & & 0.32 &  0.32 \\
$6^+_{\beta}$    & 0.31  &             & 0.30  &             & 0.32  &             & 0.30 & & 0.34 &  0.34 \\
$8^+_{\beta}$    & 0.31  &             & 0.31  &             & 0.32  &             & 0.31 & & 0.34 &  0.34 \\
$10^+_{\beta}$   & 0.31  &             & 0.31  &             & 0.32  &             & 0.32 & & 0.35 &  0.34 \\
\hline \hline
\end{tabular}
\caption{Gyromagnetic factors in $^{168-178}$Yb nuclei in units of the nuclear magneton, $[\mu_N]$. The column on the left gives the angular momentum $J$, parity $\pi$ and the corresponding band; and from second to eleventh column the experimental data \cite{Bag18,Sin95,Bro99,Bas06} and pseudo-SU(3) model calculations for $g(J^\pi_{band})$ are shown.}
\label{g-fact}\end{table}

The spectroscopic quadrupole moments that were calculated within the framework of our model are listed in Table \ref{q-moments}. These were calculated following the guidelines presented in Ref. \cite{Var17}. It is interesting to compare the existing experimental values with the ones calculated with our model. For the $2^+_{g.s.b.}$ in the $^{170-176}$Yb states, the values are described very well by the model, even within the experimental error bars. For the $4^+_{g.s.b.}$ in $^{172}$Yb, the calculated value is also within the error bars, and only the $4^+_{g.s.b.}$ in $^{174}$Yb is overestimated by the experimental data by 0.09 eb. For the other calculated states there are no experimental values to make the comparison.

When analyzing the behavior of the quadrupole moment within each band, it is observed that, in general terms, its absolute value increases. For example for the $^{168}$Yb, the $|Q|$ is $2.31~eb$ for state  $2^+_{g.s.b.}$ and rises to $3.06~eb$ in the $10^+_{g.s.b.}$ state. For the $^{178}$Yb the increase in value from $|Q|$ is even larger, going from $2.39~eb$ for the state $2^+_{g.s.b.}$ up to $3.39~eb$ in $10^+_{g.s.b.}$. In other rotational bands studied, we also observed an increase of the absolute value of the  quadrupole moment, in some cases with important increases of $2.16~eb$ units for the $\beta$-band in $^{170}$Yb. Nevertheless, there are some exceptions to this behavior, like the one presented for the $\beta$-band in $^{174}$Yb, where the $|Q|$ has a slight increase and becomes small for the state $10_\beta^+$, showing a change in collectivity within the band. 

A fact that is important to underline is that within the predictions of the model, watching the $|Q|$ as function of the number of neutrons, the maximum value is for the $\gamma$-band in $^{174}$Yb. This is the isotope of Yb with $N = 104$, so the model adequately reproduces the expected behavior of maximum collectivity right at the midshell.

\begin{table}\hspace{-2.7cm}
\begin{tabular}{c|c|cc|cc|cc|cc|c}\hline \hline
 & \multicolumn{10}{c}{$Q(J^\pi_{band})$ [eb]} \\ 
$J_{band}^{\pi}$ & $^{168}$Yb & \multicolumn{2}{c|}{$^{170}$Yb} & \multicolumn{2}{c|}{$^{172}$Yb} & \multicolumn{2}{c|}{$^{174}$Yb} & \multicolumn{2}{c|}{$^{176}$Yb} & $^{178}$Yb \\
                                                & Theo. & Exp.& Theo. & Exp. &Theo. & Exp. & Theo.& Exp. & Theo.& Theo. \\ \hline
$2^+_{g.s.b.}$     &   -2.31 &$2.1\pm0.4$ &-2.43&$2.16\pm0.37$&-2.44&$2.12\pm0.25$&-2.43 &$2.2\pm0.4$& -2.38 & -2.39 \\
$4^+_{g.s.b.}$     &   -2.92 & &-3.09 &$-2.3\pm1.2$& -3.11 &$-1.8\pm1.2$& -3.09 & & -3.01 & -3.02 \\
$6^+_{g.s.b.}$     &   -3.17 & &-3.39 & & -3.41 & & -3.40 & & -3.25 & -3.26 \\
$8^+_{g.s.b.}$     &   -3.28 & &-3.56 & & -3.58 & & -3.58 & & -3.35 & -3.36 \\
$10^+_{g.s.b.}$    &   -3.06 & &-3.67 & & -3.68 & & -3.69 & & -3.37 & -3.39 \\ \hline
$2^+_{\gamma}$  &    2.22 & & 1.04 & &  2.38 & & -1.24 & &  2.38 &  2.39 \\
$3^+_{\gamma}$  &    0.10 & &-0.09 & & -0.01 & & -2.13 & & -0.01 & -0.01 \\
$4^+_{\gamma}$  &   -1.01 & &-1.97 & & -1.25 & & -2.66 & & -1.25 & -1.25 \\
$5^+_{\gamma}$  &   -1.73 & &-2.15 & & -1.97 & & -2.95 & & -1.93 & -1.94 \\
$6^+_{\gamma}$  &   -2.32 & &-2.88 & & -2.46 & & -3.18 & & -2.45 & -2.46 \\
$7^+_{\gamma}$  &   -2.61 & &-2.99 & & -2.77 & & -3.32 & & -2.69 & -2.71 \\
$8^+_{\gamma}$  &   -2.85 & &-3.29 & & -3.03 & & -3.44 & & -3.02 & -3.04 \\
$9^+_{\gamma}$  &   -3.11 & &-3.39 & & -3.20 & & -3.52 & & -3.08 & -3.09 \\
$10^+_{\gamma}$ &   -2.68 & &-3.45 & & -3.31 & & -3.59 & & -3.30 & -3.32 \\ \hline
$2^+_{\beta}$   &   -2.03 & &-1.03 & & -2.38 & & -1.95 & & -0.57 & -0.55 \\
$4^+_{\beta}$   &   -2.67 & &-2.25 & & -3.00 & & -2.26 & &  0.76 &  0.79 \\
$6^+_{\beta}$   &   -2.93 & &-2.79 & & -3.25 & & -2.13 & &  1.20 &  1.22 \\
$8^+_{\beta}$   &   -3.15 & &-3.05 & & -3.36 & & -1.76 & &  1.44 &  1.45 \\
$10^+_{\beta}$  &   -2.90 & &-3.19 & & -3.40 & & -1.29 & &  1.59 &  1.61 \\
\hline \hline
\end{tabular}
\caption{Quadrupole moments in $^{168-178}$Yb in units of [eb]. The column on the left gives the angular momentum ($J$), parity ($\pi$) and the corresponding band; and the columns from second to eleventh are the experimental (Exp.) and pseudo-SU(3) model calculations (Theo.) for $Q(J^\pi_\alpha)$.}
\label{q-moments}\end{table}

\section{Conclusions}
\label{conclu}

The pseudo-SU(3) shell model has been implemented for the first time in $^{168-178}$Yb isotopes with the aim of making a quantitative microscopic description of the low-energy structure in this chain. The model includes the base states with pseudospin 0 and 1, a Hamiltonian with the Nilsson and pairing terms systematically parameterized and three rotor type terms, which have allowed us to describe the rotational bands adequately. The wave function obtained from the description of the energy spectrum has been used in each state to calculate the B(E2) intra and inter-band transitions, the $g$-factors and the quadrupole moments in these nuclei. The collective behavior in the ground state and $\gamma$ rotational bands is adequately described. However, it predicts a moment of inertia deficient in the {\it $\beta$} rotational bands {\bf $\beta$} in the $^{174-178}$Yb nuclei, where the predictions of the model are overrated in energy regarding the experimental values.

Most of the inter-band B(E2) transitions found in the present work are small compared with the intra-band, showing that the wave function of the states is almost orthogonal. Nevertheless, the B(E2) inter-band transitions are very underestimated regarding the experimental data, being a great limitation of the model. For the g-factors, the model shows a tendency to increase the value as a function of the angular momentum of the rotational bands or by varying the number of neutrons. However, it fails to reproduce the precise values reported experimentally, except for the $g(2_{g.s.b.}^+)$ in $^{174}$Yb. For the quadrupole moments, the model reproduces adequately the experimental values reported, and apart from the $\beta$-band in $^{174}$Yb, shows a tendency to increase $|Q|$ according to the angular momentum for each rotational band. It is important to note that the predicted maximum value for the absolute value of quadrupole moments occurs in $^{174}$Yb, this is exactly in the middle of the neutrons shell. The reported results confirm the benefits of the model in the study of rare earth nuclei.

As an extension of the present work it is proposed to investigate the contributions of proton and neutron to the values of $g$ factors. Futhermore, the extension of the calculations to heavier nuclei, such as the Hf and W isotpes, or of Yb with $110 \leq N \leq 116$ would enable us to determine the benefits and limitations of  the model to describe the transition to regions where triaxiality appears.

No funding was received for conducting this study. The authors have no competing interests which are relevant to the content of this article to declare.

\section{Acknowledgments}

We would like to thank Dr. F. Mason Lambert for the style corrections of the manuscript.


\end{document}